\documentclass[superscriptaddress, nofootinbib, preprintnumbers, twocolumn]{revtex4-1}

\usepackage{amsmath}	
\usepackage{amsthm}		
\usepackage{amssymb}	
\usepackage{eufrak}
\usepackage{datetime}
\usepackage{graphicx}
\usepackage{xcolor}
\usepackage{verbatim}
\usepackage{bm}
\usepackage{float}
\usepackage{braket}
\usepackage[colorlinks=true, citecolor=midblue, linkcolor=midblue, urlcolor=midblue]{hyperref}
\usepackage{orcidlink}
\usepackage{setspace}
\usepackage{cleveref}
\usepackage{amsmath}	% AMS Math Package
\usepackage{amsthm}		% Theorem Formatting
\usepackage{amssymb}	% Math symbols such as \mathbb
\usepackage{eufrak}
\usepackage{datetime}
\usepackage{graphicx}
\usepackage{xcolor}
\usepackage{verbatim}
\usepackage{bm}
\usepackage{float}
\onecolumngrid  % Switch to single column (instead of \onecolumn)
\twocolumngrid  % Switch back to two columns

\usepackage[colorlinks=true, citecolor=midblue, linkcolor=midblue, urlcolor=midblue]{hyperref}

\usepackage{setspace}

\definecolor{grey}{rgb}{0.4,0.4,0.4}
\definecolor{dullmagenta}{rgb}{0.4,0,0.4}
\definecolor{darkblue}{rgb}{0,0,0.4}
\definecolor{midblue}{rgb}{0,0,0.5}
\definecolor{midred}{rgb}{0.5,0,0}
\definecolor{orange}{rgb}{1,0.5,0}
\definecolor{lightbrown}{rgb}{0.75,0.5,0.25}
\definecolor{tan}{cmyk}{0.14,0.42,0.56,0}
\definecolor{djunglegreen}{cmyk}{0.99,0,0.52,0}
\definecolor{lightgreen}{rgb}{0,1,0}
\definecolor{olivegreen}{cmyk}{0.64,0,0.95,0.40}
\definecolor{midgreen}{rgb}{0.0,0.675,0.0}
\definecolor{darkgreen}{rgb}{0,0.5,0}

\smallskipamount
\settimeformat{ampmtime}

\usepackage{float}

\begin{document}

%%%%%%%%%%%%%%%%%%%%%%%%%%%%%%%%%%%%%%%%%%%%%%%%%%%%%%%%
\title{Searching for the  $N$-naturalness tower of neutrinos}

\author{Sofia Lonardi\,\orcidlink{0009-0004-3575-0335}}\email{sofia.lonardi@campus.lmu.de}
\affiliation{Ludwig-Maximilians-Universität München, Geschwister-Scholl-Platz 1, 
80539 München, Germany}

\author{Manuel Ettengruber\,\orcidlink{0000-0001-7331-6370}}
\email{manuel-meinrad.ettengruber@ipht.fr}
\affiliation{Universite Paris-Saclay, CNRS, CEA, Institut de Physique Theorique, 91191, Gif-sur-Yvette, France
}

\author{Philipp Eller\,\orcidlink{0000-0001-6354-5209}}\email{philipp.eller@tum.de}
\affiliation{Technical University Munich (TUM), James-Franck-Strasse 1, 85748 Garching, Germany}

\date{\formatdate{\day}{\month}{\year}}

\begin{abstract}
We present the first experimental search for the $N$-naturalness tower of neutrinos using a global analysis of publicly available neutrino data. As a potential solution to the hierarchy problem, the $N$-naturalness model employs a varying Higgs mass parameter among $N$ dark sectors and encodes the required fine-tuning of the placement of the sectors in a parameter $r$. We report exclusion limits on the parameter space ($r$, $N$), for normal and inverted ordering and in case the neutrino is Majorana or Dirac. In the case of a Majorana neutrino, we can rule out $N\leq 10^4$ sectors with $r\geq 0.1$. This is particularly exciting as it shows that a gauge and gravitational unification around $M_\textrm{GUT}$, which is a benchmark scenario of the $N$-naturalness framework with $N=10^4$ without fine-tuning, is ruled out by neutrino experiments.
\end{abstract}

\maketitle
%%%%%%%%%%%%%%%%%%%%%%%%%%%%%%%%%%%%%%%%%%%%%%%%%%%%%%%%

\section{Introduction}
Neutrinos are fascinating particles. The fact that the right-handed neutrino would be sterile under the Standard Model (SM) gauge group allows for a huge number of alternative mass generation mechanisms next to the well-known Higgs mechanism. The origin of neutrino mass remains an open question, which has led to a major experimental program aimed at shedding light on its nature. The extreme smallness of the neutrino mass is itself a puzzle, and may well be the first hint that the answer lies not in any known feature of the Standard Model, but in physics beyond it.

The most common way to introduce neutrino masses is via new ultraviolet (UV) physics that leads to a Weinberg operator \cite{Weinberg:1979sa} in the low-energy effective theory (for example, via the celebrated seesaw mechanism \cite{Minkowski:1977sc, Gell-Mann:1979vob, Yanagida:1980xy, Mohapatra:1979ia, Mohapatra:2004zh}). Alternatively, one can search for the answer to the small neutrino masses in the infrared. A mechanism that has been further developed in recent years is the \textit{many mixing partners mechanism}. In this framework, the neutrino acquires a small mass through the dilution of its coupling strength via a large number of mixing partners. This mechanism first appeared in the model of large extra dimensions \cite{Arkani-Hamed:1998jmv} in which the Kaluza-Klein (KK) modes of a right-handed bulk neutrino played the role of the mixing partners for the neutrino \cite{Arkani-Hamed:1998wuz, Dvali:1999cn}. In the Dvali-Redi model with many copies of the SM \cite{Dvali:2009ne}, it was recognised that this feature is not intrinsic to extra-dimensions but can be generated by any large set of neutrino mixing partners. In \cite{Ettengruber:2022pxf}, a general treatment of this mechanism and the resulting phenomenology was presented. 

Notice that, differently from standard approaches to the neutrino mass problem, the many mixing partners mechanism relies on the existence of many degrees of freedom which can, in principle, be very light. Even as light as the neutrino itself. Of course, the presence of many extra light degrees of freedom shows characteristic signatures in neutrino experiments that make this mechanism, in principle, easily distinguishable from other UV solutions. 

The many mixing partner mechanism, has been implemented in several theories that explain, next to neutrino masses, a plethora of open problems like the hierarchy problem \cite{Arkani-Hamed:1998jmv, Dvali:2009ne, Arkani-Hamed:2016rle}, dark matter \cite{Arkani-Hamed:1998sfv, Dvali:2009fw, Friedlander:2022ttk, Anchordoqui:2022txe, Gonzalo:2022jac, Anchordoqui:2022tgp, Ettengruber:2025kzw}, and the cosmological constant problem \cite{Montero:2022prj}. The explanatory power of these theories makes them exciting candidates for experimental tests from all different perspectives. In particular, neutrino experiments have been used to search for the imprints of the KK-modes of compactified extra dimensions \cite{Machado:2011jt, Basto-Gonzalez:2012nel, Girardi:2014gna, Eller:2025lsh, Elacmaz:2025ihm, Bai:2026kdq} and of additional neutrino species \cite{Ettengruber:2024fcq}. 

One of the theories that provides a solution to the dark matter and hierarchy problem simultaneously is the $N$-naturalness model \cite{Arkani-Hamed:2016rle}, in which the many mixing partners mechanism has been implemented recently \cite{Ettengruber:2025usk}. This enables the $N$-naturalness theory also to solve the small neutrino mass puzzle and simultaneously makes it possible to use terrestrial neutrino experiments to test this theory for the first time (for cosmological considerations see \cite{Han:2018pek, Bansal:2024afn}). 

In this paper, we aim to use the current publicly available neutrino data to search for the tower of new light states that the $N$-naturalness theory predicts. To do so, we combine datasets from different neutrino experiments, like GERDA and Daya Bay, and perform a combined fit of the relevant SM and $N$-naturalness parameters. We also estimate the sensitivity of future neutrino experiments to the theory of $N$-naturalness by simulating a SM-sourced neutrino spectrum of the experiments JUNO+TAO and LEGEND and calculate the possible future exclusion limit on the $N$-naturalness parameters. 

This paper is organised as follows: In section \cref{Theory} we describe the theoretical setup, in \cref{Parametrization} the mass matrix parametrisation. In \cref{Experiments} we briefly introduce the analysed experiments and in \cref{Analysis} the analysis carried out. Our results are presented in \cref{Results} and we conclude in \cref{Conclusion}
\section{N-naturalness and Neutrinos}
\label{Theory}

Let us start with describing the $N$-naturalness theory \cite{Arkani-Hamed:2016rle} and how neutrino mass can be explained within it. The original problem addressed by this theory is the hierarchy problem, which is generated by the sensitivity of the Higgs mass to the cutoff of the SM and the required fine-tuning to realise the light Higgs boson observed at LHC. 

The starting point of $N$-naturalness is the realisation that the fine-tuning issue is a problem only if the value of the Higgs mass parameter is picked as a special one within a set of other possible parameter values. For simplicity, consider an interval from zero to one, from which one must pick a number. The question then is: what is the typical distance from that number to its nearest neighbour, or to the boundaries of the interval? On average, this would be 1/2 if just one number is chosen. If, instead, one picks $N$ numbers within this interval, the nearest neighbour lies within a distance that scales with $1/N$. 

Applied to the hierarchy problem, the reasoning is the following. Imagine that the physics cutoff that stabilises the Higgs is $\Lambda_H$ and that we have several distinct gauge sectors, all exhibiting a Higgs field. The mass parameter of the Higgs, $\mu_i$, where $i$ is the label of the different gauge sectors, varies within the interval $[-\Lambda_H^2, \Lambda_H^2]$. Now, in the case of $N$ sectors, the spacing among the different Higgs mass parameters will be of order $\sim \frac{\Lambda_H^2}{N}$. Therefore, in this setup the solution to the hierarchy problem is to ensure that the spacing among the different Higgs masses is small enough such that a sector with a small Higgs mass is natural. 

Now, as a first approximation, it is useful to assume that most naturally the $\mu$'s are distributed uniformly among the interval, leading to the expression
\begin{equation}
    \mu_i^2 = - \frac{\Lambda^2_H}{N} (2i + r) \;,
\end{equation}
where $ - N/2 \leq i \leq N/2$ and $i=0$ is our sector. The parameter $r$ is parameterizing the amount of fine-tuning required to ensure a small enough Higgs mass. In case of no fine-tuning, $r$ is equal to 1, so that the spacing between sectors is exactly $\frac{1}{N}$. Nevertheless, this might not be sufficient to realise a light enough Higgs: in this case is then possible to make $r$ smaller by hand, to ensure that the mass of the Higgs is aligned with the observations. $r$ different from 1 indicates, therefore, an additional fine-tuning. Depending on the sign of $\mu_i$, the sectors either experience symmetry breaking in the usual fashion or are broken on a much lower scale by the QCD condensate, rendering these sectors much lighter and different from the SM. The sectors with a VEV are similar to the SM, and one usually identifies the lightest sector, $i=0$, with our SM (see \cite{Batell:2025hmx} for a variation where we could exist in a heavier sector). In particular, note that only half of the sectors, the ones with a non-zero VEV, contribute to the mixing when considering the neutrino phenomenology. Nevertheless, the total number of sectors is responsible for the mass suppression, and so the parameter $N$ always refers to the total number of sectors.

With this setup, the question of why the Higgs is so light becomes the cosmological question of why the lightest sector was predominantly populated during cosmological history. Intuitively, this is possible if the reheaton that produces the particles is light enough, such that the lightest sector gets dominantly populated. In particular one finds the condition $m_{\textrm{reheaton}}\leq \Lambda_H/\sqrt{N}$. 
To realise a cosmology in which the $N$-naturalness scenario can be implemented, the original paper proposed several post-inflationary models. One possibility involves a fermionic reheaton \(S\), described by
\begin{equation}
\mathcal{L}_S \supset - \kappa S^c \sum_i l_i H_i - m_S S S^c ,
%\tag{4}
\end{equation}
while another considers a scalar reheaton \(\phi\), with
\begin{equation}
\mathcal{L}_\phi = - a \phi \sum_i |H_i|^2 - \frac{1}{2} m_\phi^2 \phi^2 .
\end{equation}
With these necessary ingredients, we are left with the last question of where one would expect the cutoff $\Lambda_H$. For theories that introduce a rather large number of new particle species, $N$, one has to keep in mind that this lowers the gravitational cutoff, $M_f$, \cite{Dvali:2007hz, Dvali:2007wp} like
\begin{equation}
    M_f = \frac{M_P}{\sqrt{N}} \;.
\end{equation}
If one wishes to preserve gauge coupling unification, the number of species must satisfy \(N \lesssim 10^{4}\). This constraint implies a cutoff scale of approximately
\[
\Lambda_H \sim 10~\text{TeV}.
\]
Another interesting choice of scales arises when \(M_f = \Lambda_H\), corresponding to a scenario in which the hierarchy problem is addressed by lowering the fundamental scale \(M_f\) to the cutoff scale of the theory. This occurs at
\[
\Lambda_H = 10^{10}~\text{TeV},
\qquad
N = 10^{16}.
\]
These are the two benchmark scenarios suggested by the $N$-naturalness theory. As we will see, the latter scenario is hard to test for neutrino experiments. Therefore, we will focus on the first one, where gauge and gravitational unification happen around the grand unified scale of $10^{16}$ GeV, and around the cutoff $\Lambda_H$, the Higgs gets stabilised by new physics like SUSY. 

Now, following \cite{Ettengruber:2025usk}, we need to extend the theory to the neutrino sector and discuss its solution to the light neutrino mass generation. The first option follows the standard line of argument for generating neutrino masses in infrared (IR) scenarios with the many mixing partners mechanism \cite{Arkani-Hamed:1998wuz, Dvali:2009ne, Ettengruber:2022pxf}. The key point is that the right-handed neutrino $\nu_R$ is a singlet under the Standard Model gauge group.
As a consequence, if each sector contains its own right-handed neutrino, the left-handed neutrinos can, in general, couple to all right-handed neutrinos, including those belonging to different sectors. In this way, the left-handed neutrino of our sector effectively mixes with $N$ right-handed neutrinos. By symmetry, the same applies to the left-handed neutrinos in all other sectors.

This leads to the Dirac mass term
\begin{equation}
    (HL)_i \lambda_{ij}\nu_{Rj}\; ,
    \label{Diracoperator}
\end{equation}
where the indices $i,j$ label the different sectors, and $\lambda_{ij}$ denotes the Yukawa coupling matrix in sector space. Since all SM-like sectors are identical up to differences in the Higgs mass parameter $\mu_i$, one expects an approximate permutation symmetry among them. This implies the Yukawa structure
\begin{equation}
 \lambda_{ij}=
 \begin{pmatrix}
 a & b & b & \dots\\
 b & a & b & \dots\\
 b & b & a & \dots\\
 \dots & \dots & \dots & \ddots
 \end{pmatrix}.
 \label{Yukawa}
\end{equation}

This form reflects the coupling among different SM-like sectors, as discussed in \cite{Dvali:2009ne}. The off-diagonal entries are constrained by perturbativity, leading to the bound
\begin{equation}
    b \leq \frac{1}{\sqrt{N}} \; .
\end{equation}

The relation between the parameters $a$ and $b$ is crucial in this work, and it will be specified further in the next sections. From a symmetry standpoint, these two parameters are of the same nature, and one does not expect them to deviate too much from one another. This motivates the expectation of democratic mixing, corresponding to $a=b$, but to allow for small deviations from perfect democracy within a given sector, we keep $a$ and $b$ as independent parameters; nevertheless, the natural expectation remains $a \approx b \sim \frac{1}{\sqrt{N}}$.

An interesting alternative scenario to the introduction of additional right-handed neutrinos is to promote the fermionic reheaton to a Majorana particle that induces a Weinberg operator of the following form
\begin{equation}
\frac{1}{m_S}(\bar{L}^ci\sigma_2 H)_i \lambda_{ij}(H i \sigma_2 L)_j \,.
\label{Weinberg}
\end{equation}
This scenario combines features of a type-I seesaw mechanism with the generic suppression of couplings in the IR frameworks. As a result, the effective neutrino mass is suppressed both by the reheaton mass $m_S$ and by the fact that the effective coupling scales as $\lambda_{ij} \sim \kappa^2 \sim 1/N$, as follows from \cref{Weinberg}.

Within the $N$-naturalness framework, this additional $1/N$ suppression in the Weinberg operator is essential. In particular, cosmological consistency requires a relatively light reheaton with mass of order $m_S \sim 100\,\mathrm{GeV}$, which by itself would be insufficient to generate the observed neutrino masses without the extra suppression from the large number of species.

In this work, we will model and probe both these scenarios, which will be denoted as $N$-naturalness Dirac (NND or $_{D}$) and $N$-naturalness Majorana (NNM or $_{M}$), respectively. The Majorana scenario is, nevertheless, the main focus, as more experiments are available for the comparison and its results are stronger. 

Starting from the Lagrangian terms \cref{Diracoperator} and \eqref{Weinberg}, it is possible to explicitly write the mass matrices $M_D$ and $M_M$ for the two cases \cite{Ettengruber:2025usk}, respectively as: 
\begin{equation}\label{mass matrix D}
   \scalebox{0.6}{$M_{D}^2=\begin{pmatrix}
           \tilde{a}^2r                   &\tilde{b}^2 \sqrt{2+r}\sqrt{r}             & \cdots    &\tilde{b}^2 \sqrt{2(\frac{N}{2}-1)+r}\sqrt{r} \\
           \tilde{b}^2\sqrt{2+r}\sqrt{r}  & \tilde{a}^2(2+r)            & \cdots    & \tilde{b}^2\sqrt{2(\frac{N}{2}-1)+r}\sqrt{2+r}\\
           \vdots              & \vdots                                   & \ddots    & \vdots                  \\
           \tilde{b}^2\sqrt{2(\frac{N}{2}-1)+r}\sqrt{r} & \tilde{b}^2\sqrt{2(\frac{N}{2}-1)+r}\sqrt{2+r}                    & \cdots    &\tilde{a}^2 2(\frac{N}{2}-1)+r
          \end{pmatrix}\frac{\Lambda_{H}^2}{N\lambda}$} 
\end{equation}
and 
\begin{equation}\label{mass matrix M}
   \scalebox{0.6}{$M_{M}=\begin{pmatrix}
           \tilde{a}^2r                   &\tilde{b}^2 \sqrt{2+r}\sqrt{r}             & \cdots    &\tilde{b}^2 \sqrt{2(\frac{N}{2}-1)+r}\sqrt{r} \\
           \tilde{b}^2\sqrt{2+r}\sqrt{r}  & \tilde{a}^2(2+r)            & \cdots    & \tilde{b}^2\sqrt{2(\frac{N}{2}-1)+r}\sqrt{2+r}\\
           \vdots              & \vdots                                   & \ddots    & \vdots                  \\
           \tilde{b}^2\sqrt{2(\frac{N}{2}-1)+r}\sqrt{r} & \tilde{b}^2\sqrt{2(\frac{N}{2}-1)+r}\sqrt{2+r}                    & \cdots    &\tilde{a}^2 2(\frac{N}{2}-1)+r
          \end{pmatrix}\frac{\Lambda_{H}^2}{N\lambda m_s}$}
\end{equation}
in which
\begin{equation}
 \begin{split}
     \tilde{a}^2=(N-1)b^2+a^2 , \qquad    \tilde{b}^2=(N-2)b^2+2ab \,,
 \end{split}
 \end{equation}
and
\begin{equation}
    \tilde{a}^2 -\tilde{b}^2=(a-b)^2.
\end{equation}
It is worth noting that from \cref{Weinberg} we get an expression which is (overall factor aside) the square of the one from \cref{Diracoperator} because of the quadratic dependence on the Higgs VEV. This allows us to make the same assumptions on the parameters and diagonalise the same matrix for both cases, using a uniform parametrisation across the two cases. 

To study the neutrino phenomenology, we need first to calculate the masses of the neutrinos in the $N$ sectors.
Diagonalising the matrices perturbatively \cite{Ettengruber:2025usk} yields the following expressions for the Dirac and Majorana neutrino masses:
\begin{equation}
m_{iD} = \sqrt{(a-b)^2\frac{\Lambda_{H}^2}{N\lambda}(2i+r)} \, ,
\end{equation}
\begin{equation}
m_{iM}=(a^2-b^2)\frac{\Lambda_{H}^2}{N\lambda m_s}(2i+r) \, .
\end{equation}
Inserting the chosen scaling for the couplings $a$ and $b$, we obtain
\begin{equation}\label{masses dirac}
m_{iD}\sim\sqrt{\frac{1}{N}\frac{\Lambda_{H}^2}{N\lambda}(2i+r)}\sim\frac{1}{\sqrt{N}}v_i \, ,
\end{equation}
\begin{equation}\label{masses majorana}
m_{iM}\sim\frac{1}{N}\frac{\Lambda_{H}^2}{N\lambda m_s}(2i+r)\sim\frac{1}{N}\frac{v_i^2}{m_S} \, .
\end{equation}

It is first of all evident that the effect of the number of sectors $N$ is suppressing all the neutrino masses of the tower equally, while the $r$ parameter has a more subtle effect, especially evident for small neutrino masses. Additionally, there exists one mode that receives a much heavier mass that scales as
\begin{equation}
    m_{HD} \sim v \sqrt{\frac{N}{2}\left(\frac{N}{2}-1+r\right)} \, ,
\end{equation}
\begin{equation}
    m_{HM} \sim \frac{v^2}{m_S} \frac{N}{2}\left(\frac{N}{2}-1+r\right)  \, .
\end{equation}

The two cases scale differently with $N$, resulting in a much stronger suppression of neutrino masses for the Majorana scenario.
In the Dirac case, $N=10^4$ is insufficient to reproduce the observed neutrino mass scale unless additional assumptions about the couplings are introduced. On the other hand, when $N=10^{16}$, the suppression reduces the required cancellation between $a$ and $b$ to about $10^{-3}$, which is compatible with the Yukawa couplings of other fermions. 
In the Majorana case, the suppression is stronger, thanks also to the additional reheaton mass appearing in the VEV expression. For $N=10^4$ the Yukawa tuning is already comparable to that of Standard Model couplings, while for $N=10^{16}$ the neutrino mass scale becomes of order $10^{-5} \text{eV}$. 
This difference between the two possible neutrino natures is also evident when considering the mass-squared differences between the neutrino in our sector and that of a generic sector $i$, which takes the form
\begin{equation}
\Delta m_{i0}^2\sim\frac{\Lambda_{H}^2}{\lambda N^2 r}i \,,
\end{equation}
for the Dirac case, and
\begin{equation}
\Delta m_{i0}^2\sim\frac{\Lambda_{H}^4}{\lambda^2 m_s^2 N^4 r}(i^2-ir) \, ,
\end{equation}
for the Majorana case. 

The diagonalisation of the mass matrix also provides the mixing angles, from which we can extract the expression for the mass eigenstate corresponding to a neutrino in our sector
\begin{equation}
\ket{\nu_0}=\ket{\nu_0}_m+ \frac{2}{N} \sum_{i=1}^{\frac{N}{2}-1}\frac{\sqrt{2i+r}\sqrt{r}}{2i} \ket{\nu_i}_m+ \frac{2}{N}\ket{\nu_N}_m  \,.
\end{equation}
From this expression, we can compute the survival probability of a neutrino produced in our sector
\begin{equation}
P(L/E)=\left|1+\frac{4}{N^2}\sum_{i=1}^{\frac{N}{2}-1}\frac{(2i+r)r}{4i^2}e^{i\Phi_i}+\frac{4}{N^2}e^{i\Phi_N}\right| \, ,
\end{equation}
where
\begin{equation}
\Phi_i=\frac{L}{2E}\Delta m_{i0}^2 .
\end{equation}

\section{Parametrisation of the mass matrix}
\label{Parametrization}

In this work, we are interested in constraining the two main parameters of the $N$-naturalness theory ($N$, number of sectors and $r$, measure of fine-tuning) using data from different neutrino experiments.
To do so, we must develop a model that makes predictions for the neutrino oscillations, the effective electron neutrino mass $m_\nu$, and the effective Majorana mass $m_{\beta\beta}$, which depend on these parameters.
We start from the mass matrices \cref{mass matrix D} and \cref{mass matrix M}: diagonalising them yields the squared masses $m_{i}^2$ of the neutrinos in each of the $N$ sectors and their corresponding eigenvectors.
These results are necessary to calculate both the oscillations and the contributions to the effective masses. From $m_{i}^2$, we can calculate the mass splittings $\Delta{m_{ij}^2}$.
The mixing matrix $V$, which diagonalises $M^2$, encodes the mixing between different sectors and can be used to calculate the oscillations and the different contributions to $m_\nu$ and $m_{\beta\beta}$.
This model is implemented in the \textit{Newtrinos.jl} framework \cite{eller_newtrinos_2025}, a framework that allows for global analysis of multiple neutrino experiments and has already been used in previous work \cite{Ettengruber:2024fcq, Kozynets:2024xgt, Eller:2025lsh, Eller:2026urd}.

To have a model that can be used in this analysis, we need to find a parametrisation of the matrix that reduces the number of free parameters, to focus on the two that we aim to constrain: the number of sectors $N$ and the fine-tuning $r$. 
Furthermore, a realistic three-flavour scenario must be taken into consideration to include the three neutrino generations in each sector. 

In \cite{Ettengruber:2025usk}, the matrices are diagonalised using a perturbative approach, in which the unperturbed term corresponds to setting $a=b$ and the first-order perturbation is proportional to the difference of the Yukawa couplings $a-b$, yielding the expressions \cref{masses dirac}, \cref{masses majorana}. This approach is an approximation, and the perturbative term becomes bigger than the unperturbed one in some regions of the parameter space of interest (especially for small $m_0$ and small $N$). 
Given these limitations, we carry out a numerical diagonalisation of the complete matrix, with a parametrisation that enables us to fix the high number of free parameters. We explain here the procedure for the Majorana case, which can be easily extended to the Dirac case, just with a different overall scale and taking the square root of the matrix.
We start by normalising the complete mass matrix by dividing all terms by $\tilde{b}^2$

\begin{equation}
   \scalebox{0.6}{$M_{M}=\begin{pmatrix}
           \eta r                   &\sqrt{2+r}\sqrt{r}             & \cdots    & \sqrt{2(\frac{N}{2}-1)+r}\sqrt{r} \\
           \sqrt{2+r}\sqrt{r}  &\eta(2+r)            & \cdots    &\sqrt{2(\frac{N}{2}-1)+r}\sqrt{2+r}\\
           \vdots              & \vdots                                   & \ddots    & \vdots                  \\
           \sqrt{2(\frac{N}{2}-1)+r}\sqrt{r} & \sqrt{2(\frac{N}{2}-1)+r}\sqrt{2+r}                    & \cdots    &\eta(2(\frac{N}{2}-1)+r)
          \end{pmatrix}\frac{\tilde{b}^2 \Lambda_{H}^2}{N\lambda m_s}$}
\end{equation}

In this way, the parameter $\tilde{b}^2$ is absorbed into the overall scale, while a new parameter $\eta=\frac{\tilde{a}^2}{\tilde{b}^2}$ appears as a multiplicative factor on diagonal elements.
For a three-flavour scenario, we would have three matrices of this kind and therefore three distinct values for $\tilde{b}^2$ and $\eta$.
Moreover, we have three more free parameters ($\Lambda_H$, $\lambda$ and $m_s$ in the Majorana case), so in total 9 values that we would like to fix. 

The three $\eta$ values must be specified: this is conceptually analogous to the perturbation approach assumption that $(\tilde{a}^2-\tilde{b}^2)\approx\frac{1}{N}$ because setting $\eta$ corresponds to fixing the relation between $a$ and $b$, the intra- and inter-sector couplings.
No theoretical basis exists for this relation to exhibit flavour dependence; therefore, all three $\eta$ values should possess identical scaling, reducing the number of independent parameters from three to one.
Concerning this one remaining $\eta$, multiple assumptions are viable, depending on the theoretical interpretation and analysis objectives.
From the Lagrangian mass term structure, no reasons exist for $\tilde{a}^2$ and $\tilde{b}^2$ to differ substantially. This symmetry-based argument corresponds to the perturbation approach's assumption of their small difference.
A strong hierarchy between a and b is also possible, such that $\tilde{b}^2\ll\tilde{a}^2$, reflecting significantly weaker inter-sector interactions, but introducing this hierarchy by hand asks for a theoretical reason why the inter-sector couplings should be so much weaker.
As a choice for this analysis, we take the natural one $\eta=1+\frac{1}{N}$, equivalent to $\tilde{a}^2-\tilde{b}^2=\frac{\tilde{b}^2}{N}$, which therefore follows \cite{Ettengruber:2025usk}. Small deviations of $\mathcal{O}(1)$ from this regime do not affect our conclusions, but of course, if we turn down the inter-sector coupling by hand, many effects would vanish. 

The next step is to fix the three overall scales and therefore the three values of $\tilde{b}^2$: this can be done using oscillation measurements.
In fact, in our sector-0, there exist three different neutrino masses $m_{1}$, $m_{2}$, $m_{3}$, which, depending on the mass ordering, correspond to the different generations. They can be expressed, using this parametrization, as
\begin{equation}
   m_1=(\eta-1 )\tilde{b_1}^2\frac{\Lambda_{H}^2}{N\lambda m_s}r \,,
\end{equation}
\begin{equation}
   m_2=(\eta-1 )\tilde{b_2}^2\frac{\Lambda_{H}^2}{N\lambda m_s}r \,,
\end{equation}
\begin{equation}
   m_3= (\eta-1 )\tilde{b_3}^2\frac{\Lambda_{H}^2}{N\lambda m_s}r \,.
\end{equation}

These expressions confirm that $\tilde{b^2}$ is a flavour-dependent parameter and therefore the three different versions need to be fixed separately using the three different mass scales. Moreover, being $\tilde{b^2}$ a multiplicative factor together with $\frac{\Lambda_{H}^2}{N\lambda m_s}$, we can treat all these parameters as an overall scale $s$ that can be fixed using the neutrino masses
\begin{equation}
   s_1= \tilde{b_1}^2\frac{\Lambda_{H}^2}{N\lambda m_s}=\frac{m_1}{(\eta-1 )r} \, ,
\end{equation}
\begin{equation}
   s_2= \tilde{b_2}^2\frac{\Lambda_{H}^2}{N\lambda m_s}=\frac{m_2}{2(\eta-1 )r} \,,
\end{equation}
\begin{equation}
   s_3= \tilde{b_3}^2\frac{\Lambda_{H}^2}{N\lambda m_s}=\frac{m_3}{(\eta-1 )r} \,.
\end{equation}
The three mass values are not independent: they are related through two measured parameters, $\Delta{m_{21}}^2=7.53\cdot10^{-5}$ eV$^2$ and $\Delta{m_{31}}^2=2.4 \cdot 10^{-3}$ eV$^2$.
Once we decide on Normal or Inverted mass ordering, $m_1$ or $m_3$ will correspond to $m_0$, respectively, and the other two masses can be calculated using the measured mass splittings.

Therefore, we are left with only one last free parameter, $m_0$: this value corresponds to the absolute mass scale of neutrinos, and experimentally we have only upper limits on it. Therefore, in this analysis, we explore different $m_0$ values ([0, $10^{-4}$, $10^{-2}$, $10^{-1}$] eV) to study the effects on the bounds in \cref{Results} and, in the final results, we fix it to $m_0=0.01$ eV, which corresponds to a conservative and experimentally justified assumption.
In this way, all the 9 free parameters have been fixed or rewritten in terms of $N$ or $m_0$, leaving the mass matrix dependent only on $N$ and $r$, exactly the parameters to constrain.

Employing this parametrisation and generalising it to a three-flavour framework, three distinct matrices require diagonalisation and subsequent combination to obtain the complete mixing matrix.
The three matrices $M_M$ are diagonalized separately, to facilitate the tracking of each eigenvalue's generation. This produces three distinct sets of eigenvalues and eigenvectors requiring reorganization into a $\frac{3N}{2} \times \frac{3N}{2}$ configuration and subsequent multiplication by $U_{PMNS}$ to obtain $V_{final}$.
We can, in this way, derive the eigenvalues with the appropriate associated eigenvectors, encoding the full mixing.
The next step is calculating the predictions of the model in terms of observables such as survival probabilities and effective masses.
For the oscillations, we need to derive the mass splittings as differences between each neutrino mass and the minimum mass. This is done by generalising the Standard Three-Flavour oscillations calculation.
In fact, the PMNS matrix must be generalised to an $N$-sector scenario. The full mixing matrix becomes $U^{\frac{3N}{2} \times \frac{3N}{2}}_{PMNS}V$, where $U$ is an expanded version of the PMNS matrix in $\frac{3N}{2} \times \frac{3N}{2}$ dimensions, and $V$ is the eigenvector matrix from diagonalisation, encoding mixing among different sectors.
As an example, the survival probability of a neutrino flavour $l$ within our sector 0 is expressed as
\begin{equation}
\scalebox{0.8}{$
P(\nu_{0l}\rightarrow \nu_{0l}) \approx \sum^{N}_{i,j=0}  \sum_{h,k=e,\mu,\tau}W_{lk0i}W_{lh0j}^{*}W_{lh0j}W_{lk0i}^{*} e^{-i\frac{\Delta m_{hkij}^2L}{2E}}$} \, ,
\end{equation}
where $W_{lk0i}=U_{lk}V_{0i}$.
Using the mass splittings and the mixing angles, oscillation probabilities are calculated.
The mass matrix eigenvalues also enable the calculation of the effective mass from $\beta$-decay, that is, the mass of an electron antineutrino emitted with an electron. In the Standard Three-Flavour scenario, this is given by
\begin{equation}\label{me SM}
    m_{\nu}^2= \sum_{i=1}^3  |U_{ei}|^2 m_i^2 \,.
\end{equation}
In  $N$-naturalness, we have $N$ eigenvalues per flavour and the new mixing matrix $W$. 
We thus generalise this to
\begin{equation}\label{me NN}
    m_{\nu}^2= \sum_{i=1}^3\sum_{j=0}^{\frac{N}{2}} |U_{ei}|^2|V_{0j}|^2 m_{ij}^2 \,.
\end{equation}
Similarly, eigenvalues and eigenvectors allow the determination of the effective Majorana mass of a virtual Majorana neutrino participating in neutrinoless double-$\beta$ decay.
In the Standard Three-flavour scenario, this is
\begin{equation}\label{mb SM}
m_{\beta\beta}=|\sum_{i=1}^3  U_{ei}^2 m_i| \,.
\end{equation}
In  $N$-naturalness, we generalise this as
\begin{equation}\label{mb NN}
m_{\beta\beta}=|\sum_{i=1}^3\sum_{j=0}^{\frac{N}{2}}  (U_{ei}V_{0j})^2 m_{ij}| \,.
\end{equation}
The described parametrisation, therefore, allows the calculation of theoretical predictions of $N$-naturalness for the different observables. Through their comparison with experimental data, in the next sections, we can set bounds on the free parameters of the model.

\section{Neutrino experiments}
\label{Experiments}
In this analysis, we make use of publicly available data and future projections from terrestrial neutrino experiments, which measure either neutrino oscillations or neutrino effective mass.

In the final results, we present contours from Daya Bay, JUNO+TAO, GERDA, LEGEND-200 and LEGEND-1000. Other experiments, such as MINOS and KamLand  have been taken into consideration, but since they are not sensitive enough to the theory parameters, at least at the analysis level performed, they are not reported here.

\subsection{Oscillation experiments}
We make use of two reactor neutrino experiments: Daya Bay and JUNO+TAO.

The Daya Bay Reactor Neutrino Experiment was a short-baseline ($\sim$ 1.65 km) reactor experiment located near Shenzhen, China, composed of 8 antineutrino detectors split across two near halls and one far hall. The reactors produced a flux of antineutrinos of 2.9 GW in the energy range of [1-10] MeV, and this was measured via inverse-beta-decay from both the near and far detectors, to highlight the eventual lack of antineutrinos, and therefore detect electron antineutrino oscillations \cite{an_detector_2016, collaboration_side-by-side_2012}. The ratio between its baseline and the energy of the neutrino is optimal to estimate precisely the oscillation parameters $m^2_{31}$ and $\theta_{13}$. Publicly available data and the forward model of the experiment refer to \cite{daya_bay_collaboration_precision_2023}.

JUNO, the Jiangmen Underground Neutrino Observatory, is a next–generation reactor antineutrino experiment located in Guangdong, China, about 52.5 km from the Taishan and Yangjiang nuclear power plants. The main goal of the experiment is to measure electron antineutrino oscillations and determine the neutrino mass ordering. The energy of neutrinos is of the order of a few MeV, and the longer baseline allows the experiment to be very sensitive to both solar and atmospheric oscillation parameters ($m^2_{21}$ and $\theta_{12}$, $m^2_{31}$ and $\theta_{13}$). 
The detectors contain a 20 kton target of liquid scintillator enclosed within a 35.4 m diameter acrylic sphere surrounded by 17612 large 20-inch photomultiplier tubes (PMTs) and immersed in a water pool filled with ultra-pure Cherenkov water, which serves as a cosmic muon veto \cite{abusleme_potential_2025, abusleme_sub-percent_2022}. TAO is the near detector, and its main purpose is to provide a reference spectrum for JUNO, eliminating the possible model dependence due to fine structure in the reactor antineutrino spectrum. The energy resolution of the detector is 3\% at around 1 MeV.
The experiment started operations in August 2025 and published its first results in June 2026 \cite{JUNO2026}. At the moment of writing, therefore, only a few months of data are available, and for this reason, in this work, we only show projections for a six-year exposure.

\subsection{Neutrinoless double $\beta$-decay experiments}
GERDA (GERmanium Detector Array) is an experiment located at the INFN Laboratori Nazionali del Gran Sasso (LNGS) designed to search for neutrinoless double $\beta$-decay ($0\nu\beta\beta$) of $^{76}$Ge \cite{gerda_collaboration_final_2020}. 
In the Standard Model $2\nu\beta\beta$ decay, two neutrinos are emitted together with two electrons. In contrast, in the $0\nu\beta\beta$ process, the neutrino is exchanged virtually between the two decay vertices and no neutrinos appear in the final state. As a consequence, the entire decay energy is carried by the two emitted electrons, producing a sharp peak at the characteristic $Q_{\beta\beta}$ energy of the isotope.
GERDA employs high-purity germanium (HPGe) detectors enriched in $^{76}$Ge. These detectors are arranged in seven strings placed inside a cryostat filled with liquid argon (LAr), which serves both as a cooling medium and as an active shielding system. The cryostat is further surrounded by a large water tank containing ultra-pure water instrumented with photomultiplier tubes, which acts as passive shielding and as a Cherenkov detector for the residual cosmic muons reaching the underground laboratory.
GERDA did not see any signal and has set a limit $T^{0\nu}_{1/2} \geq1.8\times10^{26}$ yrs \cite{gerda_collaboration_final_2020}.

The LEGEND project builds upon the experience gained from GERDA. Its goal is to perform a next-generation search for neutrinoless double $\beta$-decay with substantially improved sensitivity \cite{Saleh:2025bdy} given mainly by the augmented detector mass (200 and 1000 kg for LEGEND-200 and LEGEND-1000, respectively). LEGEND-200 collaboration recently released the first results from the run started in 2023 and, combined with the result of GERDA, it provides a limit on the half-life time of $T^{0\nu}_{1/2} \geq1.9\times10^{26}$ yrs \cite{Saleh:2025bdy}. Since the result is still very close to GERDA's one, we only show projections for the expected bound at the end of operation ($T^{0\nu}_{1/2} \geq 10^{27}$ yrs).

\section{Analysis}
\label{Analysis}
We use publicly available data from each neutrino experiment to constrain the two main parameters of $N$-naturalness that affect neutrino phenomenology: $N$ and $r$. Using a profiled likelihood-ratio test statistic and Wilk's theorem we draw exclusion contours in the parameter space, performing first each analysis separately, and subsequently combining the different experiments.

\subsection{Oscillation analysis}
\begin{figure}[h]
\centering
\includegraphics[width=\linewidth]{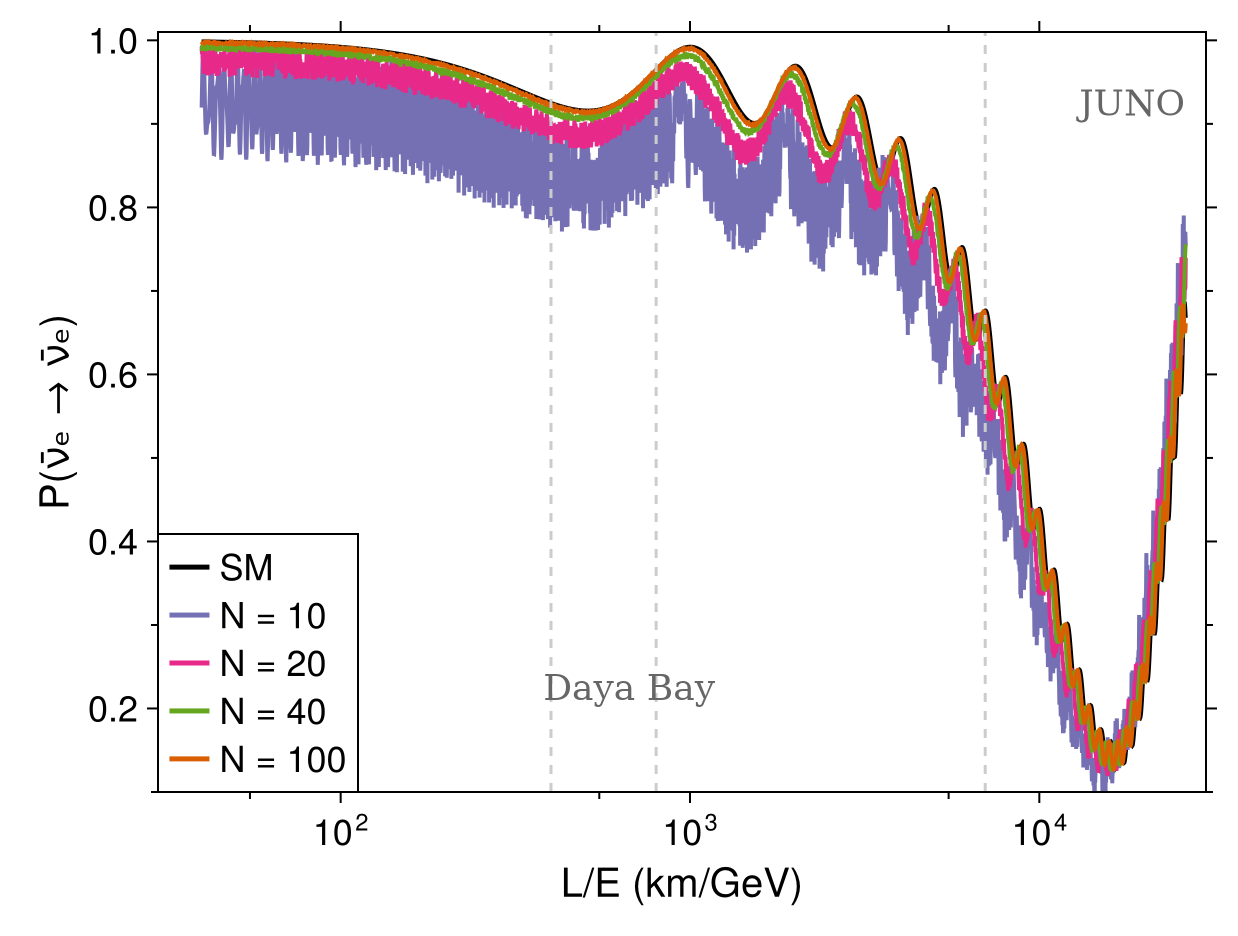}
\caption{Survival probability of electron antineutrino for different values of $N$, for Majorana NO with $r=1$, $m_0$=0.01 eV. }
\label{fig dayabay osc N}
\end{figure}

\begin{figure}[h]
\centering
\includegraphics[width=\linewidth]{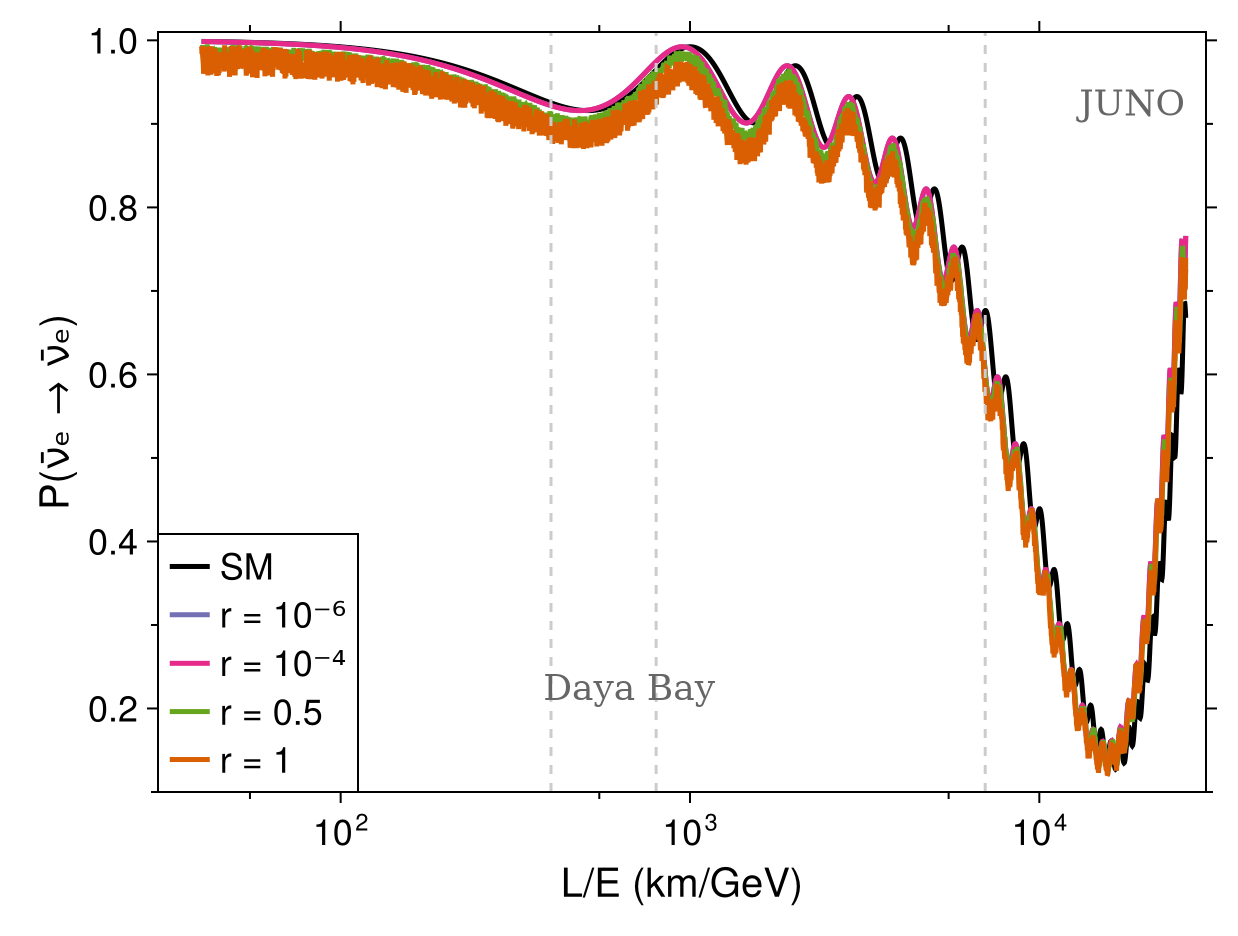}
\caption{Survival probability of electron antineutrino for different values of $r$, for Majorana NO with $N=20$, $m_0$=0.01 eV. }
\label{fig dayabay osc r}
\end{figure}

To illustrate how the $N$-naturalness modifies neutrino oscillations, we show oscillograms in the mode and energy range to which the two experiments, Daya Bay and JUNO, are sensitive, i.e. ([1-40] MeV and 1.65 km) and ([0.5-8.5] MeV and 52.5 km) for Daya Bay and JUNO, respectively.
We compute the survival probability of electron antineutrinos for different values of the parameters $N$ and $r$, separately for the Dirac and Majorana cases, and for Normal and Inverted ordering (NO and IO).
In \cref{fig dayabay osc N,fig dayabay osc r}, the case of Majorana Normal ordering is shown as an example.

This is representative of the effect of the many sectors on the oscillations: one can clearly see the fast oscillations generated by the heavy modes. We can analyze the main effects of the parameters on oscillations: $N$ acts as an overall suppression, making the minimum of the oscillation probability deeper for small values of $N$, while the effect of $r$ is more subtle, mainly affecting the energy of the oscillation minimum. As $N$ increases, the predictions gradually approach the Standard Three-Flavour oscillations (SM), whose phenomenology is recalled in the limit of $N\rightarrow \infty$ and  $r\rightarrow0$. It is worth noticing that the $r$ effect is more evident in the Daya Bay region than in JUNO's: this is because $r$ mainly affects the very heavy modes and JUNO, due to its baseline, is less sensitive to them, as will also become evident in the final results.
A similar phenomenology also emerges in the Inverted Ordering and Dirac cases, with the difference that the Majorana case has enhanced modifications due to the quadratic dependency on the Higgs vev and therefore a stronger sensitivity to the parameters than the Dirac case.   

The effects of the parameter $m_0$ and $\eta$ on the oscillations are also studied: the first has a similar effect to $N$, and it recalls the Standard oscillations in the limit $m_0\rightarrow0$, while the second washes out the effect of $r$ for high values. Since these two parameters correspond to a physical choice of the mass scale and the ratio of the couplings, if not differently specified, they are fixed to $m_0=0.01$ eV and $\eta=1+\frac{1}{N}$, as we are testing this specific realisation of the model.  

\begin{figure}[h]
\centering
\includegraphics[width=\linewidth]{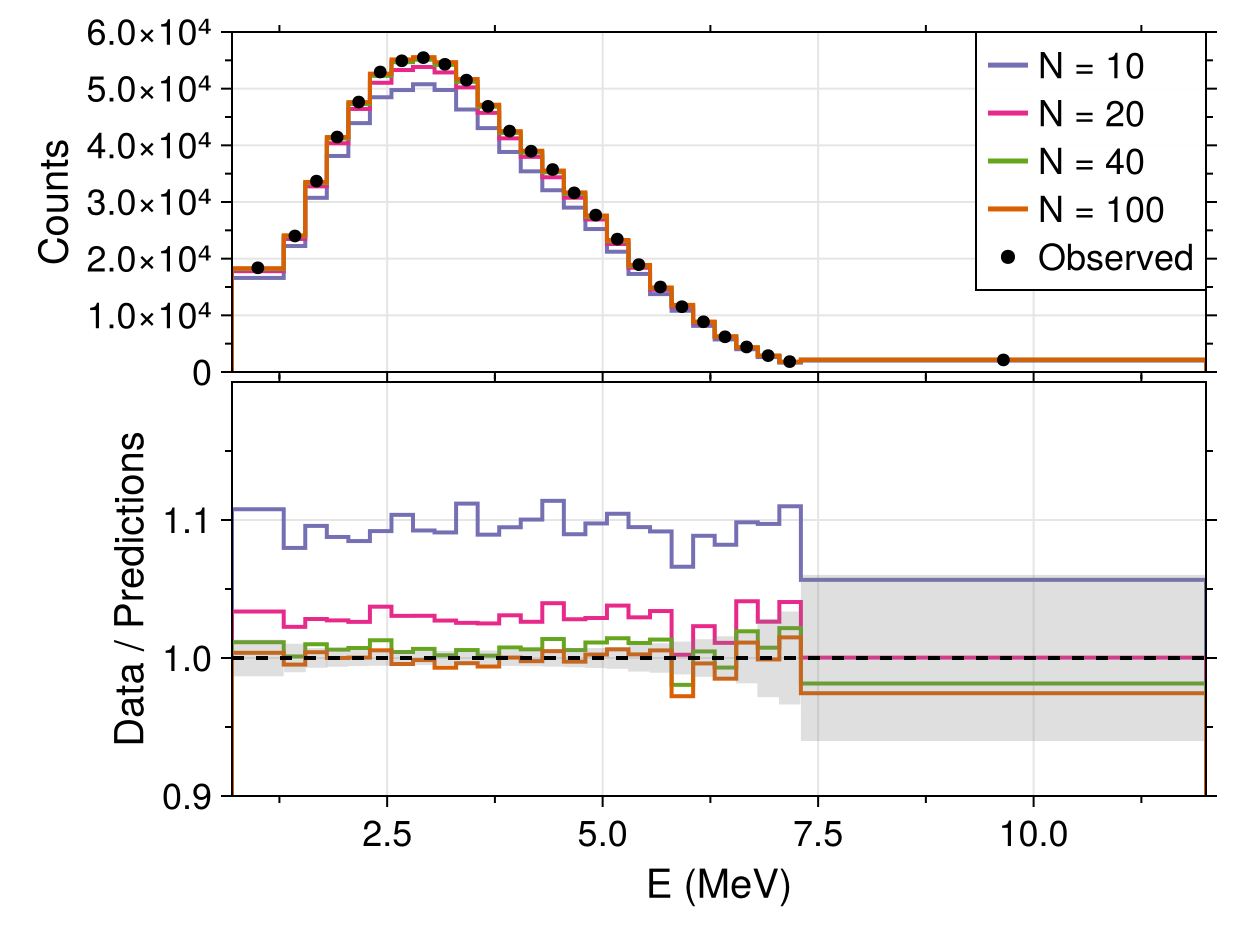}
\caption{Counts of Daya Bay data and predictions for different values of $N$, in the case Majorana NO with $r=1$, $m_0$=0.01 eV.}
\label{fig dayabay data N}
\end{figure}

\begin{figure}[h]
\centering
\includegraphics[width=\linewidth]{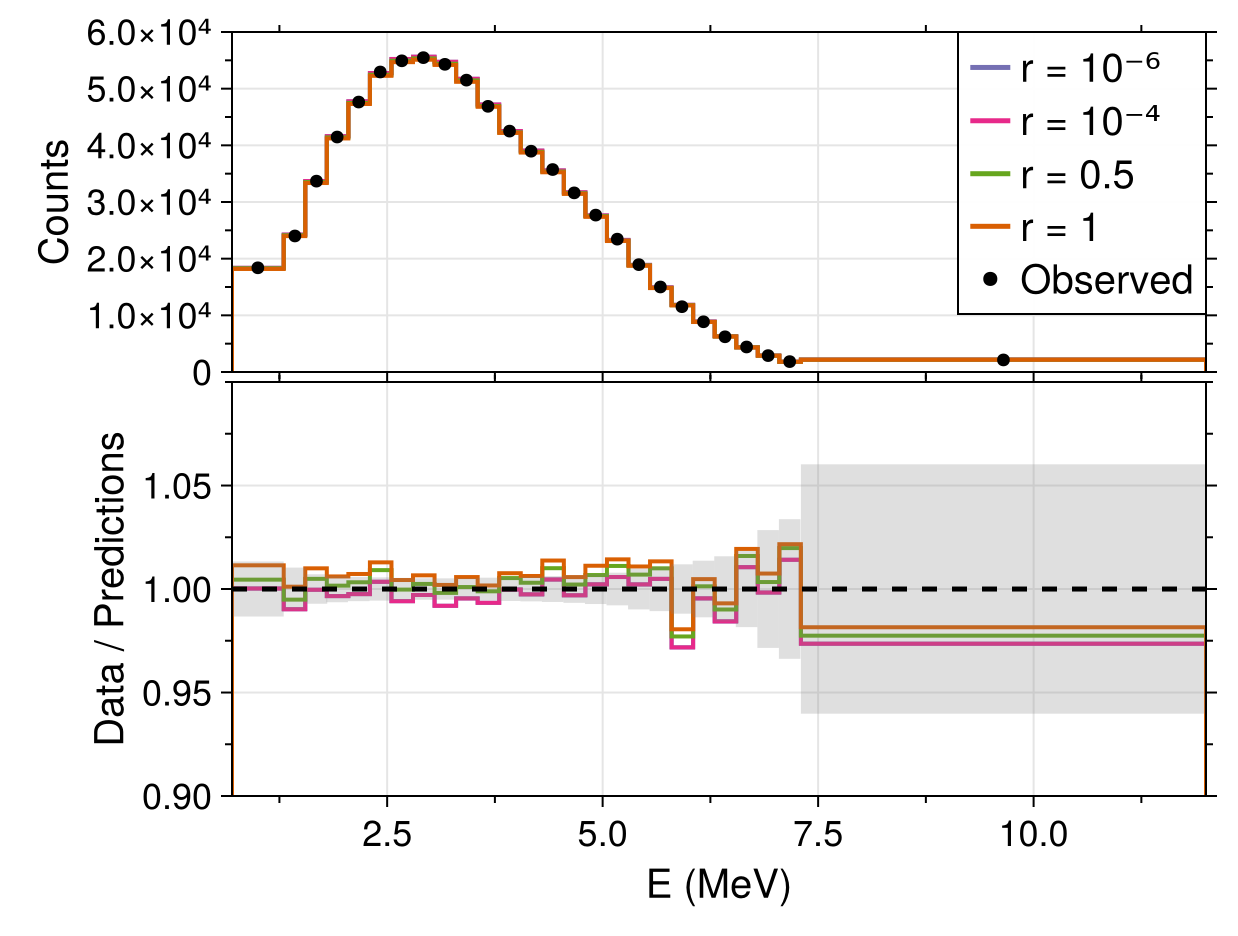}
\caption{Counts of Daya Bay data and predictions for different values of $r$, in the case Majorana NO with $N=20$, $m_0$=0.01 eV. }
\label{fig dayabay data r}
\end{figure}

To illustrate the effect on an experiment's observables, we compare Daya Bay public data with the $N$-naturalness-predicted counts in the Majorana NO case for different values of $N$ and $r$ in \cref{fig dayabay data N,fig dayabay data r}.


To compare data and predictions, we use a profile likelihood ratio test statistic $t(\vec{\theta})$, the ratio between the likelihood evaluated at each point in a grid vs. the likelihood at the best-fit parameters $\vec{\theta}^{\star}$
\[
\lambda(\vec\theta) =
\frac{L(\vec\theta)}
     {L(\vec\theta^\star)} \,,
\]
\[
t(\vec\theta) = -2 \log \lambda(\vec\theta)\,.
\]
Using Wilks' theorem, the test statistic $t(\vec{\theta})$ can be assumed to follow a $\chi^2$ distribution asymptotically, with degrees of freedom equal to the number of parameters. This allows us to compute excluded regions at different confidence levels. In this analysis, we draw contours with a 90\% confidence level.
Two-dimensional scans are performed over the free parameters $r$ and $N$, while the remaining parameters (nuisance parameters), arising from the oscillation framework and the experimental implementation (such as flux normalisation and energy smearing), must be treated separately.
The parameters $m_0$ and $\eta$ are fixed, if not differently specified, to test one specific realisation of the model. 
The neutrino oscillation parameters are fixed to values compatible with global analyses of neutrino data \cite{esteban_nufit6_0_2024, Capozzi:2025wyn, ParticleDataGroup:2024cfk}.
The additional parameters appearing in the implementation of the forward models are profiled over in the likelihood calculation, as they reflect the uncertainty on the modelisation of the experimental setup and could impact the results. In the Daya Bay model, there are no additional parameters; systematics are handled via a covariance matrix, while in JUNO+TAO, there are several of them adjusting the energy scale and resolution, and the flux normalization. JUNO and TAO together have 22 nuisance parameters over which we profiled.

\subsection{Neutrinoless double $\beta$-decay analysis}

Besides the information provided by oscillation experiments, the absolute scale of neutrino masses cannot be inferred from oscillations alone, since they depend only on mass squared differences. The presence of additional sectors in the model affects the Majorana effective mass estimated in the search for neutrinoless-double-$\beta$-decay $m_{\beta\beta}$.
The idea is therefore to compare the expected value of this observable in the $N$-naturalness with the best fit found by the experiments.
This approach uses the data collected by neutrinoless-double-$\beta$-decay experiments only indirectly: the Majorana effective mass $m_{\beta\beta}$ is calculated from the inverse of the half-life time
\begin{equation}\label{eq:gerda_half_life}
 \frac{1}{T^{0\nu}_{1/2}}= G \cdot M_N \cdot \frac{m_{\beta\beta}^2}{m_e^2} \,,
\end{equation}
where $G$ is the phase-space factor and $M_N$ is the nuclear matrix element.
The experiments can extract only lower limits on $T^{0\nu}_{1/2}$ given the lack of a signal peak in the energy spectrum, and therefore, there is no way to have a direct comparison between the prediction from $N$-naturalness and the actual data from the experiment.
For these reasons, we decided to work on a point-like comparison between the mass estimations only. This study applies only to the Majorana scenario since this is the only option for neutrinoless-double-$\beta$-decay.

We start from $m_{\beta\beta}$ prediction of $N$-naturalness in \cref{mb NN} and we compute the corresponding $T^{0\nu}_{1/2}$ using  \cref{eq:gerda_half_life} with this value of the effective mass. We can then compare the obtained values with the measured and estimated upper limits
\[
T^{0\nu}_{1/2} \geq
1.8\times10^{26},
\;
10^{27},
\;
10^{28}
\ \text{yrs} \,,
\]
at 90\% C.L. of GERDA, LEGEND-200 and LEGEND-1000 respectively \cite{gerda_collaboration_final_2020, Saleh:2025bdy, acharya_first_2026}. 

A major source of uncertainty in this calculation is the nuclear matrix element $M_N$. Significant differences arise among the various theoretical approaches used to compute it \cite{Engel2017NME, Ettengruber:2022mtm, pompa_impact_2023}.  
Furthermore, $M_N$ depends on the mass of the exchanged neutrino. Since heavy neutrinos may appear in the $N$-naturalness scenario, their contribution must be properly included. We follow the treatment proposed in \cite{Bolton:2022hnl}, where the effective mass is written as
\begin{equation}
m_{\beta\beta}^{\mathrm{eff}}
=
\left|\sum_{i}^{N_A} U_{ei}^{2} m_i
+
\sum_{k}^{N_S} U_{eN_k}^{2} m_{N_k}
\frac{\langle p^2 \rangle \mathcal{F}(m_{N_k})}{\langle p^2 \rangle + m_{N_k}^{2}}
\right| \,,
\end{equation}
where $N_A$ denotes the light neutrino states ($m_\nu \leq 10^4$ eV) and $N_S$ the heavier states ($m_\nu \geq 10^4$ eV). The momentum $\langle p^2 \rangle$ is given by
\begin{equation}
\langle p^2 \rangle = m_e m_p \left| \frac{\mathcal{M}^{0\nu}_N}{\mathcal{M}^{0\nu}_\nu} \right| \,.
\end{equation}
in which $\mathcal{M}^{0\nu}_\nu$ and $\mathcal{M}^{0\nu}_N$ are the matrix elements for light and heavy neutrinos, respectively.
The interpolation correction function $\mathcal{F}(m_{N_k})$ is taken to be 0.7 for $10^4 \leq m_\nu \leq 10^6$ eV and 1 for $m_\nu \geq 10^6$ eV.
For $^{76}$Ge the nuclear matrix elements span the ranges
\[
\mathcal{M}^{0\nu}_\nu \in [2.89, 6.04], \qquad
\mathcal{M}^{0\nu}_N \in [104, 401].
\]
Following \cite{Bolton:2022hnl}, we adopt the benchmark values
\[
\mathcal{M}^{0\nu}_\nu = 5.28,
\qquad
\mathcal{M}^{0\nu}_N = 194 \,.
\]
Studies are made on the final contours to check if they are strongly affected by this choice, and the results vary only slightly, making this assumption justified.

We are interested in the effect of the parameters $N$ and $r$ for the effective mass estimation, but in this case also in the parameter $m_0$, as it sets the overall scale of the neutrino tower. As an example, we take the result of GERDA and compare it with the $N$-naturalness Majorana NO predictions.

\begin{figure}[h]
\centering
\includegraphics[width=\linewidth]{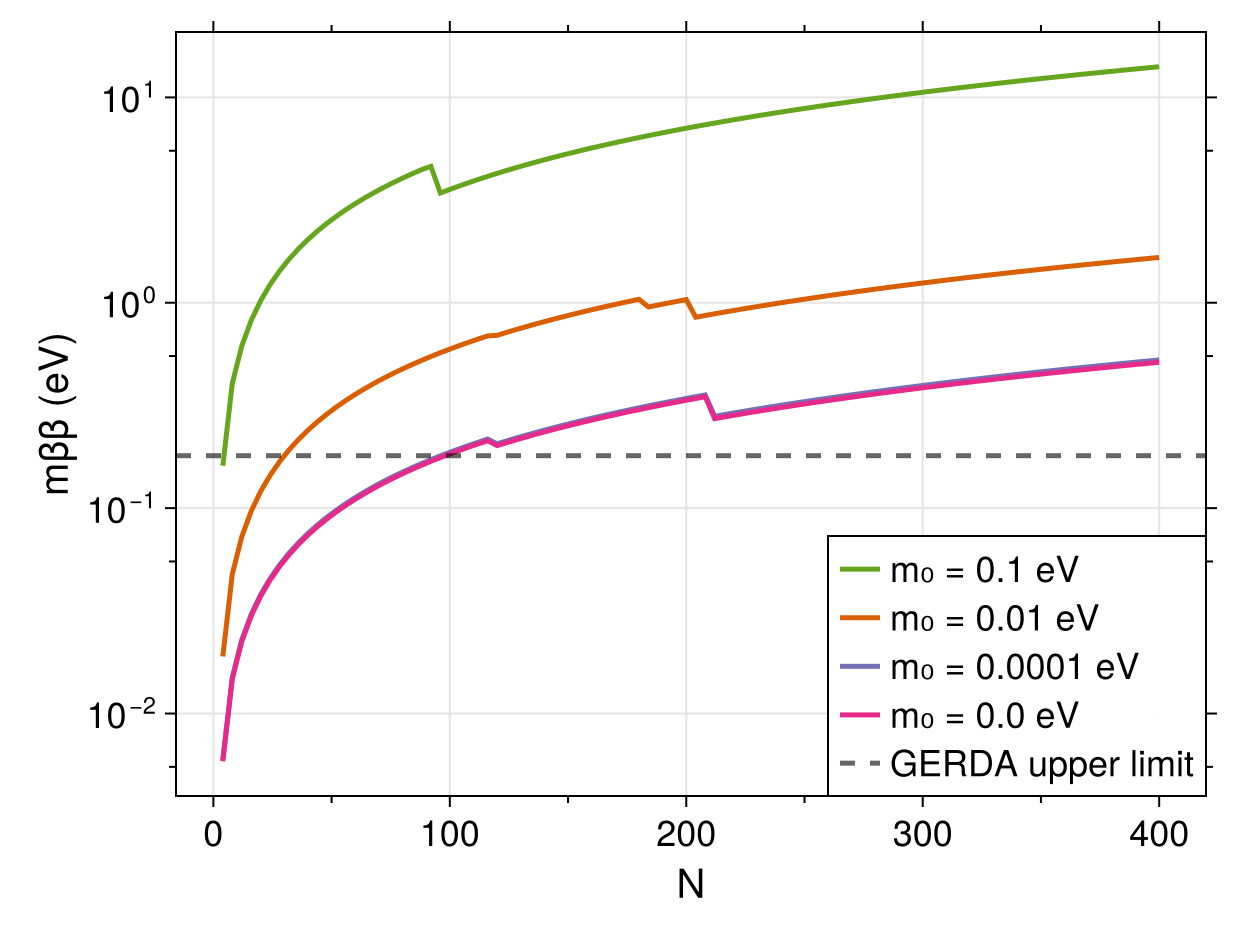}
\caption{Effective mass $m_{\beta\beta}$ as a function of the number of sectors $N$ for Majorana NO for different values of $m_0$.}
\label{fig:gerda_N_NO}
\end{figure}
\begin{figure}[h]
\centering
\includegraphics[width=\linewidth]{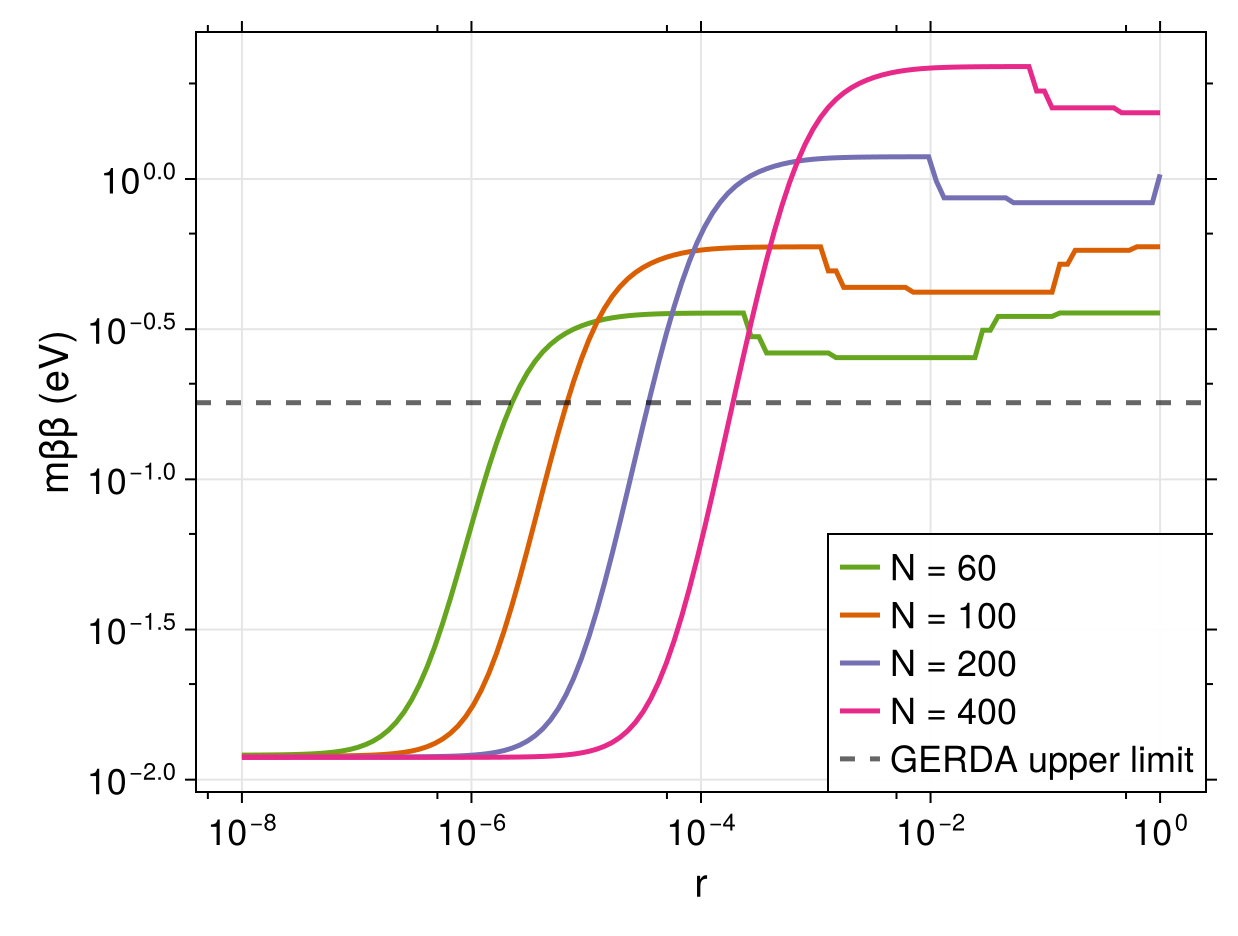}
\caption{Effective mass $m_{\beta\beta}$ as a function of the fine-tuning parameter $r$ for Majorana NO for different values of $N$.}
\label{fig:gerda_r_NO}
\end{figure}

From \cref{fig:gerda_N_NO} we observe that $m_{\beta\beta}$ increases with the number of sectors $N$ and also grows with the lightest neutrino mass $m_0$. 
Small discontinuities appear when neutrino masses reach the MeV scale due to the transition between the light and heavy nuclear matrix element treatments. 

\Cref{fig:gerda_r_NO} illustrates the dependence on the parameter $r$. For each value of $N$ there exists a region where the predicted effective mass is reduced because of the treatment of the matrix elements.  
Overall, comparing these predictions with the values inferred from GERDA measurements indicates that the experiment should be sensitive to both $N$ and $r$, allowing meaningful constraints to be derived.
The likelihood analysis is performed as previously explained for the oscillations experiments, but comparing with the prediction only one value of $T^{0\nu}_{1/2}$ with an estimated uncertainty of $\sim 10\%$.

\section{Results}
\label{Results}
We scan the parameter space for every experiment, and we combine all of them in order to derive the final constraints, reflecting both current and projected bounds on the different realisations of the theory.

Rather than simply overlaying the excluded regions obtained from individual experiments, we perform a combined analysis that consistently incorporates the information from all datasets. 
This is achieved by combining the likelihood functions of the individual experiments, from which we determine the joint best-fit region and the corresponding exclusion limits, following the procedure described previously.

As a final result, we produce two comprehensive scans, corresponding to the Majorana realisations of the $N$-naturalness framework in Normal and Inverted ordering. These scans represent the main results. We also present the contours for the Dirac case, in both Normal and Inverted ordering, as a comparison, even though they do not exclude a big portion of the parameter space and are, in this sense, weaker.
Each scan shows both the excluded regions derived from the individual experiments and the constraints obtained from the global fit. 
It is important to note that these results are obtained for a fixed value of $m_0$ = 0.01 $\mathrm{eV}$ and $\eta = 1 + \frac{1}{N}$.
In particular, the value of $m_0$ is chosen conservatively based on current cosmological limits, while its impact on the effective mass predictions is studied separately. The parameter $\eta$ corresponds to the one suggested by naturalness as argued above.
Under these assumptions, we can restrict the scans to the parameters $r$ and $N$.
In the figures, the grey shaded regions correspond to the parameter space excluded by current data, while the black dashed lines indicate the projected exclusion limits expected after six years of future data.

\begin{figure}[h]
\centering
\includegraphics[width=\linewidth]{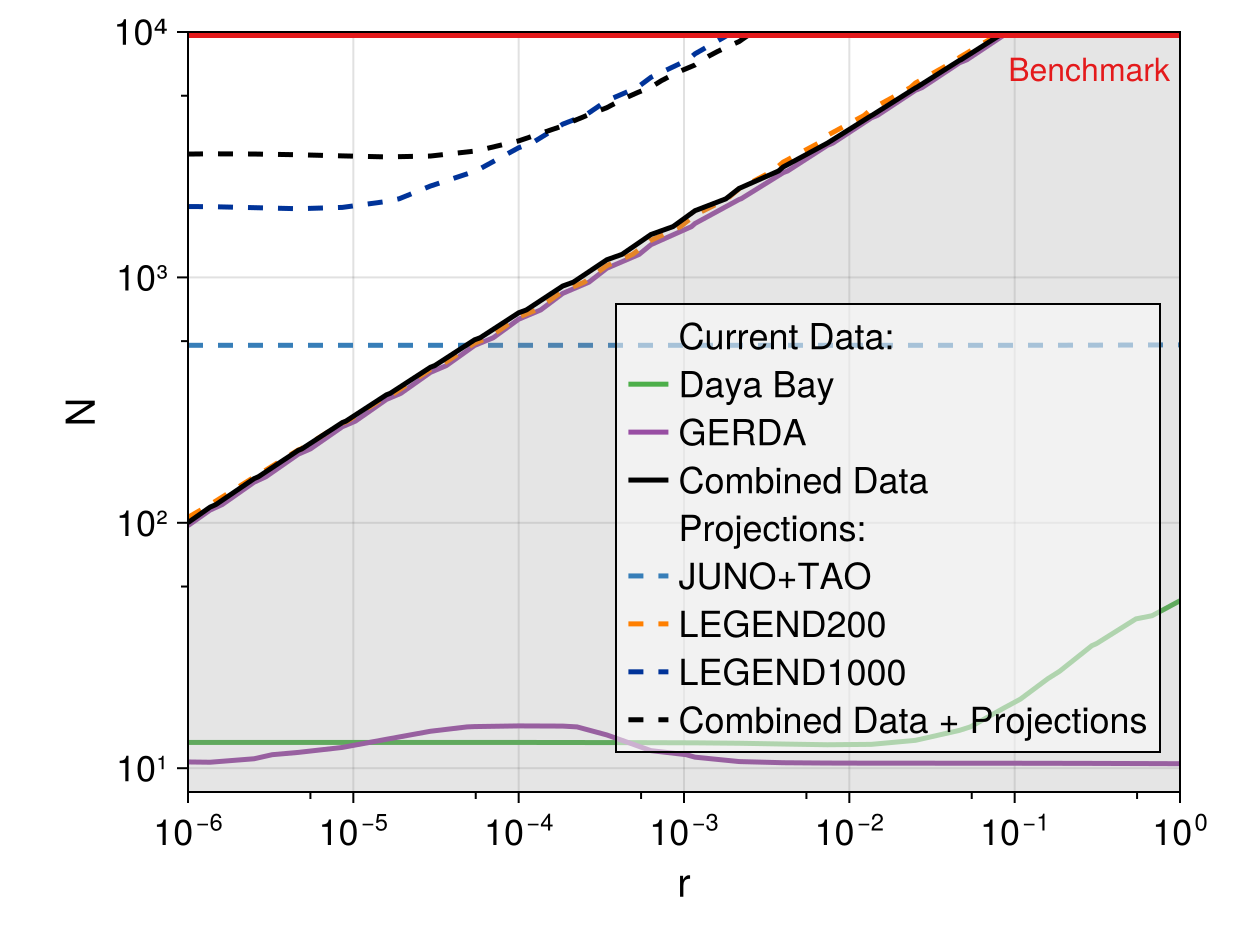}
\caption{Excluded regions at 90\% C.L. in the $(r, N)$ plane from all the experiments considered for the Majorana NO model of  $N$-naturalness. Solid lines are bounds given by current data, while dashed lines are simulated data for 6 years. The filled grey areas correspond to currently excluded regions. }
\label{fig maj no}
\end{figure}

\begin{figure}[h]
\centering
\includegraphics[width=\linewidth]{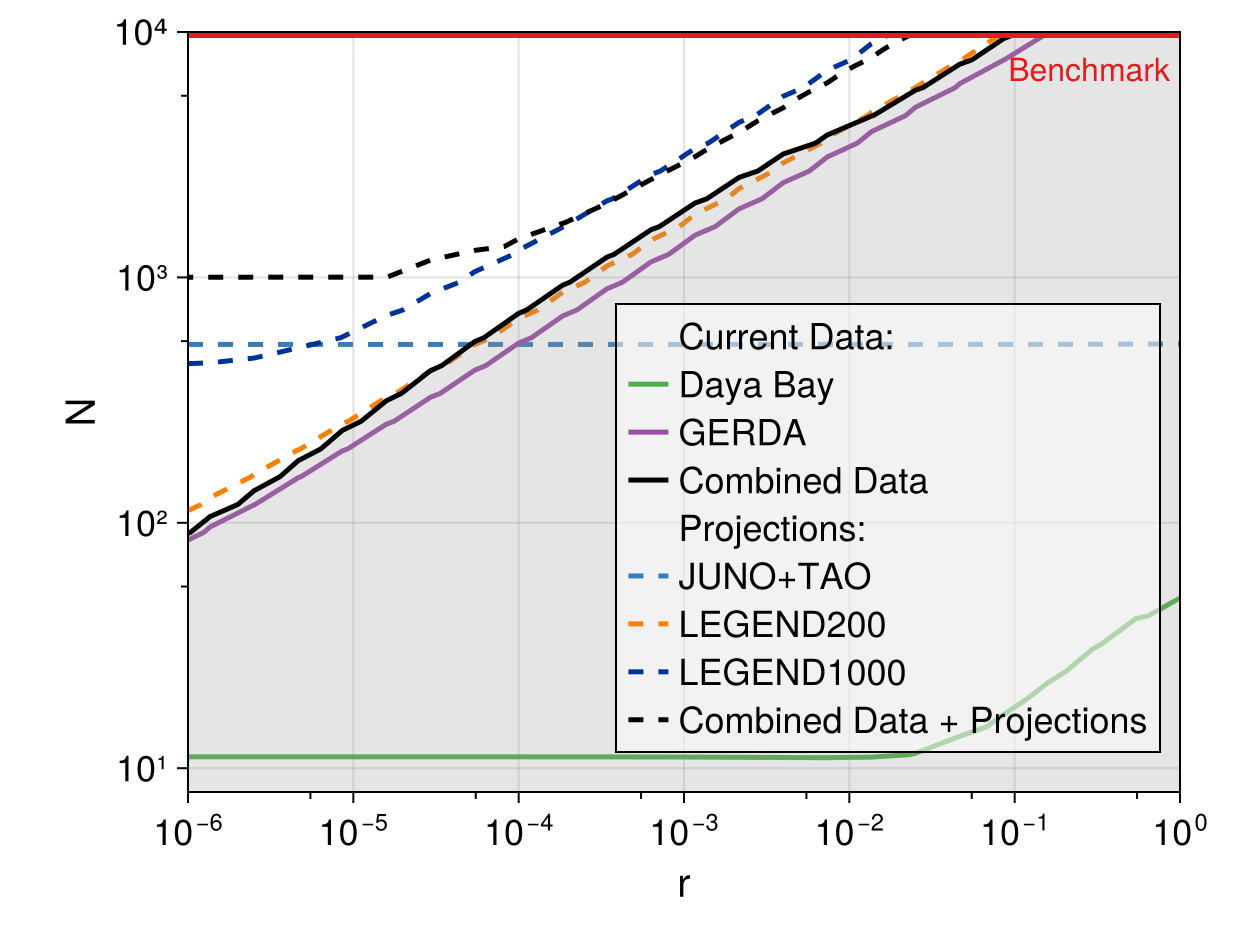}
\caption{Excluded regions at 90\% C.L. in the $(r, N)$ plane from all the experiments considered for the Majorana IO model of  $N$-naturalness. Solid lines are bounds given by current data, while dashed lines are simulated data for 6 years. The filled grey areas correspond to currently excluded regions. }
\label{fig maj io}
\end{figure}

The strongest constraints are obtained for the Majorana realisations, shown for Normal and Inverted Ordering in \cref{fig maj no,fig maj io}, respectively.

The contours for Normal and Inverted Ordering are very similar, and the sensitivity is dominated by the GERDA experiment. 

With the current GERDA data combined with Daya Bay, it is already possible to exclude all non-fine-tuned scenarios with $r = 1$ up to $N = 10^4$. Increasingly strong fine-tuning is required as $N$ decreases, reaching values as small as $r \sim 10^{-6}$ for the minimal allowed value of $N \sim 100$. Considering the JUNO+TAO projected sensitivity and the corresponding global constraints, all values of $N < 500$ are excluded for any choice of $r$. In the region $500 \leq N \leq 10^4$, the theory remains viable only if a fine-tuning of at least $r\leq 0.1$ is introduced.

\begin{figure}[h]
\centering
\includegraphics[width=\linewidth]{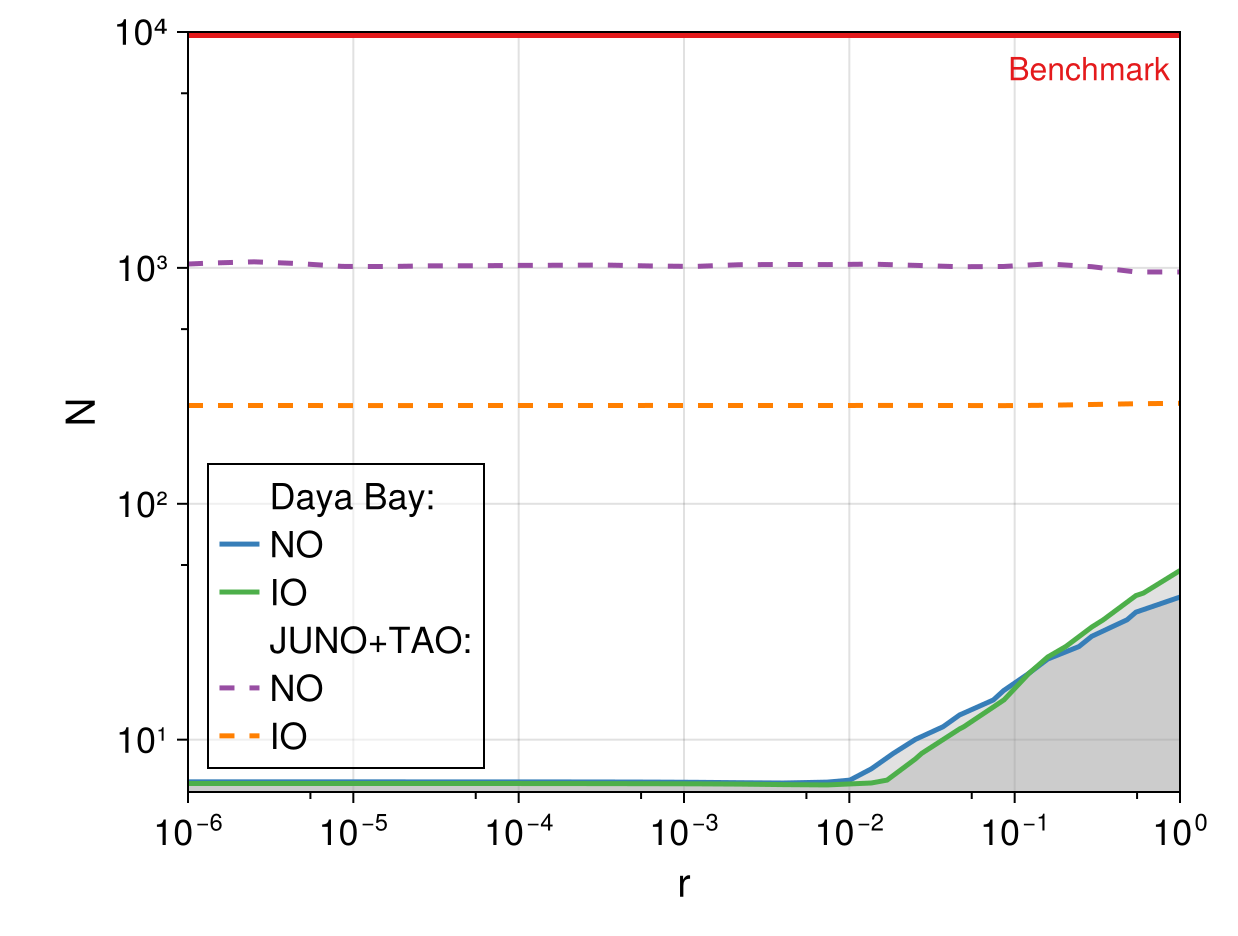}
\caption{Excluded regions at 90\% C.L. in the $(r, N)$ plane from all the experiments considered for the Dirac model NO and IO of $N$-naturalness. Solid lines are bounds given by current data, while dashed lines are simulated data for 6 years. The filled grey areas correspond to currently excluded regions. }
\label{fig dir no}
\end{figure}

The situation changes drastically in the Dirac case, shown for both Normal Ordering and Inverted Ordering in \cref{fig dir no}. 
The Dirac case is the least constrained. This is mainly because neutrinoless double-$\beta$-decay experiments are not applicable. Daya Bay excludes the small $N$ region ($N \leq 50$) for values of $r\sim1$ (non-fine-tuned realisations). As a result, the non-fine-tuned scenarios ($r=1$) are excluded by current data only up to $N$ of order 50.
The projected sensitivity from JUNO+TAO, indicated by the dashed contours, will extend the exclusion region to $N$ values greater than $10^2$, thereby narrowing the gap. However, JUNO is less sensitive to the parameter $r$, which limits its ability to exclude non-fine-tuned realisations of the model.

While the contours given to the Dirac model are limited, it is worth focusing on the Majorana case: these results indicate that non-fine-tuned realisations of the $N$-naturalness theory are already excluded for $N \lesssim 10^4$. 
This is particularly interesting because it directly probes one of the model's theoretical benchmarks. As discussed in \cref{Theory}, values around $N=10^4$ could simultaneously address the hierarchy problem (assuming a GUT-scale cutoff) and generate suppressed neutrino masses, without introducing ad hoc small Yukawa couplings and preserving the gauge coupling unification.
For the Majorana case, this realisation without fine-tuning is already excluded by current data for both Normal and Inverted Ordering. 
\begin{figure}[h]
    \centering
    \includegraphics[width=\linewidth]{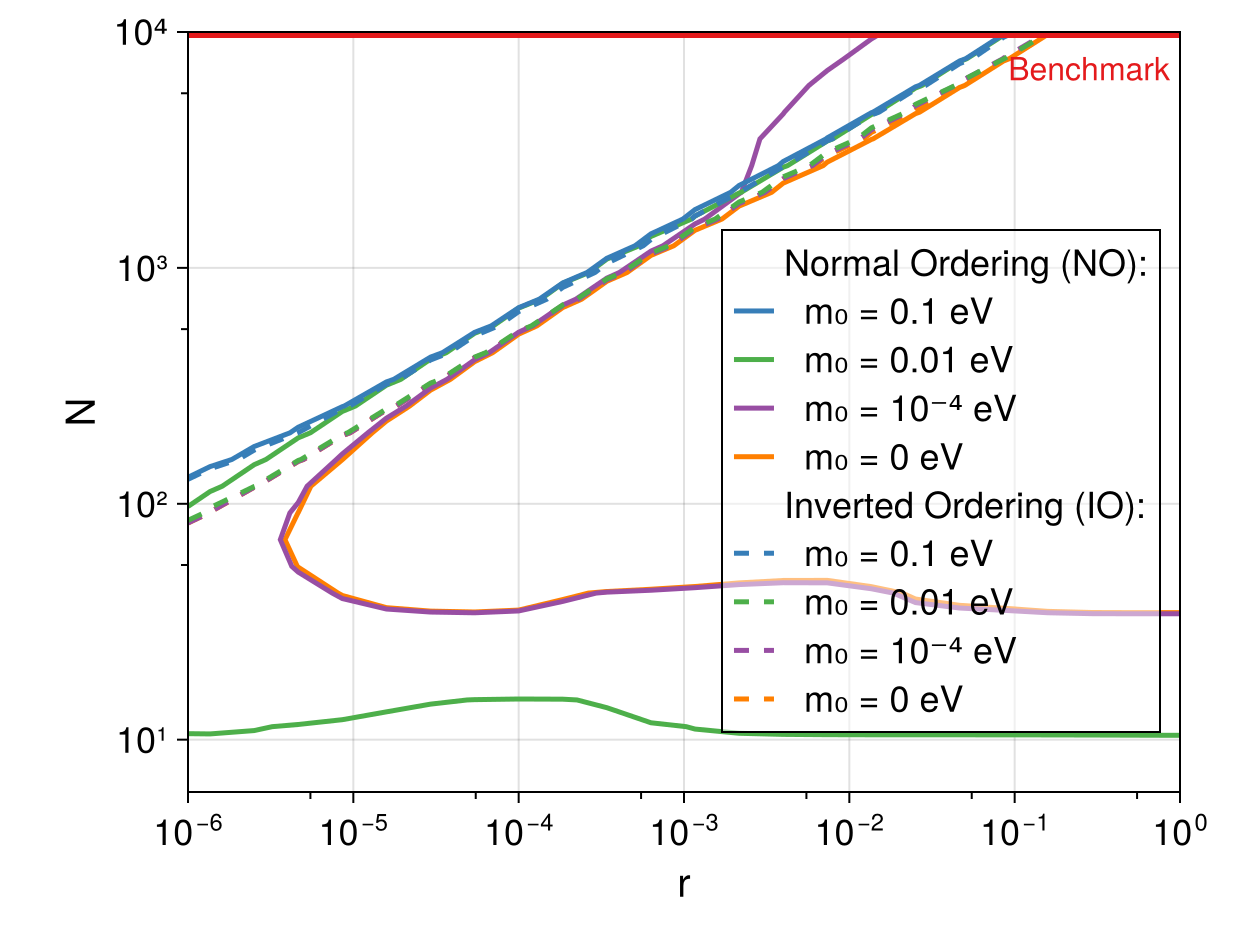}
  \caption{Excluded regions at 90\% C.L. in the $(r, N)$  plane from GERDA for the Majorana NO and IO models. These results do not vary significantly in the region of interest close to $r=1$ when varying $m_0$.}
     \label{fig ger 5}
\end{figure}
It is also interesting to see that these results remain true across all values of $m_0$, from zero to the terrestrial bound $m_0=0.1$ eV. 
In \cref{fig ger 5}, it is shown how the contours given by GERDA (the experiment giving the stronger contours) do not change significantly in the region of interest (high $N$ and $r$ close to 1) when assuming a different $m_0$, both for Normal and Inverted Ordering.

\section{Conclusion}
\label{Conclusion}

This work provides, to the best of the authors' knowledge, the first constraints on the $N$-naturalness theory from terrestrial experiments.

To obtain these results, we studied the $N$-naturalness theory and its implications for the neutrino sector, starting from the parametrisation of the mass matrix for both Dirac and Majorana cases. The phenomenology of the different $N$-naturalness models was investigated, focusing on the predictions for neutrino oscillations and neutrino effective mass estimates and their dependence on the parameters $N$, the number of sectors, and $r$, the measure of fine-tuning.
These predictions were then compared to public or simulated data from neutrino experiments, in particular Daya Bay, JUNO+TAO, GERDA, and LEGEND, to extract the regions of parameter space excluded by current observations or by projected experimental sensitivities. The entire analysis was carried out using the \textit{Newtrinos.jl} framework.

The obtained results show that neutrino experiments have the potential to test a large portion of the parameter space of the $N$-naturalness theory. In particular, we establish lower bounds on $N$ of order $10^2$ for all models using JUNO+TAO projections, and upper bounds on $r$ of order $10^{-6}$ for $N\sim100$ using GERDA, in the Majorana case. These results exclude, in the Majorana case, the non-fine-tuned model realisations ($r=1$ up to $r=0.1$), thereby invalidating the central requirement of the theory to solve the hierarchy problem without additional fine-tuning. The benchmark scenario of the $N$-naturalness with $N=10^4$ without fine-tuning ($r=1$) is ruled out by current data for the Majorana case in both Normal and Inverted Ordering.

There are several ways to improve this analysis. 
As discussed in \cref{Parametrization}, also $m_\nu$, the effective electron antineutrino mass, is sensitive to the $N$-naturalness modification, therefore including the KATRIN experiment would be particularly valuable. Heavy neutrinos would produce kinks and distortions in the spectrum, and so performing a full spectrum analysis could lead to stronger bounds, especially interesting for improving the contours in the Dirac case.
Furthermore, additional future experimental projects could be included, especially those with high sensitivity to mass splittings. 

Nevertheless, this work demonstrates the potential of the neutrino sector to constrain N-naturalness scenarios and highlights the power of using and combining public data within a unified analysis framework to exploit the complementarity of different experiments. The bounds presented here represent the first step in the exploration of the parameter space of the N-naturalness theory. 

\section*{Acknoledgements}
The authors would like to thank Alan Zander for useful discussions.
The work of S.L. and P.E. has been supported by the Deutsche Forschungsgemeinschaft (DFG, German Research Foundation) under Germany’s Excellence Strategy—EXC2094/2–390783311, and the Grant No. SFB 1258–283604770.  The work of M.E. was supported by ANR grant ANR-23-CE31-0024
EUHiggs.

\setlength{\bibsep}{5pt}

\bibliographystyle{utphys}
\bibliography{refs}

\end{document}